# Thermodynamic Properties of NiAs-FeN Phase from First Principles

Alexey KARTSEV[1,2,*] and Nina BONDARENKO[3]

[1]Research Tomsk State University, 36, Lenina pr., Tomsk, 634050, Russia

[2]Service de Physique de lEtat Condens DSM/IRAMIS/SPEC (CNRS UMR 3680), CEA Saclay, 91191 Gif-sur-Yvette Cedex, France

[3]Condensed Matter Theory Group, Physics Department, Uppsala University, S-751 21 Uppsala, Sweden

*Corresponding author

**Abstract.** We present a study of the thermodynamic properties of ferromagnetic FeN phase with nickel arsenide structure in the frameworks of the density functional perturbation theory and generalized gradient approximation (GGA). Discussion of the thermodynamic properties of FeN is given with respect to effect of external hydrostatic pressure applying. The pressure-induced enhancement of Debye temperature is predicted. The vibration densities of states at different pressures calculated by DFPT approach show the phonon band gap opening at pressure 50 GPa.

**Introduction**

Modern developments of technologies give a challenge to improve the high-temperature performance of technological materials such as metal nitrides. Therefore, new materials based on the transition metal mononitrides are of great interest for research both from industrial and scientific points of view due to their exceptional properties, ranging from superconductivity, hardness to high melting temperature and corrosion stability. Their structural, electronic properties are based on interplay between different mechanisms on the microscopic scale. Traditional phenomenological methods have reached their limits of applicability for these systems. Thus, the further improving of a theoretical description of the nitride materials needs to be based on more advanced methods, such as *ab initio* calculations of the finite-temperature thermodynamic properties. In this work, we will employ quasiharmonic approximation (QHA) in the frameworks of density functional perturbation theory (DFPT). The method is one of most widely used approaches to new material design with the inclusion of finite temperature effects on the desired properties [1].

Most of the transition metal nitrides known up to day are formed by nonmagnetic metals and tend to crystallize in the rocksalt and zinc blende structure, however the crystalline structure of equiatomic magnetic *3d* metal mononitrides is still under debates [2,3,4]. In the previous systematic work, the electronic, structural and dynamic properties of noval ferromagnetic (FM) iron mononitride (FeN) phase with nickel arsenide (NiAs) structure were presented with respect to external pressure and it has been proposed that the phase is dynamic stable [5]. FM NiAs-FeN potentially has a vast number of technological applications, and in the same time, can play an important role in addressing the issue of Earth innercore structure [6]. The aim of this work is to explore in detail vibrational and thermodynamic properties of FM NiAs-FeN as a function of pressure and temperature. Using DFPT calculations based on QHA we have evaluated free energy, the temperature dependence of Debye temperature, specific heat and phonon density of states (DOS) for NiAs-FeN phase. The rest of the paper is organized as follows. In Section II we briefly describe methods and approaches used in this work. In Section III we discuss and present results of our first principle calculations. In Section IV we summarize conclusion of this paper.

**Computational Method and Formalism**

First principle density functional theory (DFT) calculations are performed using the Quantum Espresso simulation package, based on pseudopotential plane-wave method [7]. Exchange-correlation effects were treated by utilizing of the Perdew-Burke-Ernzerhof (PBE) generalized gradient approximation [8], and a cutoff energies are required to expand the electronic wave functions and charge density set up to 60 Ryd and 600 Ryd respectively. For electronic structure calculation Brillouin zone is sampled using reciprocal grid of 24×24×14 according to Marzari-Vanderbilt cold smearing method [9] with 0.272 eV smearing. Atoms are relaxed within Hellman-Feynman forces criterion less then 0.02 eV/Å. Phonon modes are obtained on the groundwork of DFPT [10] using 6×6×6 $q$-points mesh within frequency convergence less then 0.5 cm$^{-1}$.

The theoretical analysis for thermodynamic properties presented here is based on QHA [11]. With this approximation, the Helmholtz free energy and Debye temperature at specific temperature $T$ can be determined by the vibrational spectrum via the standard harmonic expression:

$$F_{tot}(T,V) = E_0(V) + F_{vib}(T,V) + F_{el}(T,V) + F_{mag}(T,V), \qquad (1)$$

where $E_0$ is the ground state total energy, $V$ the volume, $F_{vib}$ the phonon contribution to the free energy, $F_{el}$ the thermal excitations energy and $F_{mag}$ is magnetic contribution. Within QHA vibration part can be written as harmonic expression

$$F_{vib}(T,V) = \int_0^\infty d\omega g(\omega,V) \cdot \left[ \frac{\hbar\omega}{2} + k_B T \cdot \ln\left\langle 1 - \exp\left(\frac{-\hbar\omega}{k_B T}\right)\right\rangle \right], \qquad (2)$$

The phonon DOS $g(\omega,V)$ has to be calculated. For Fermi-Dirac distributed electrons the energy of thermal excitations can be approximated like

$$F_{el}(T,V) = -\sum_{\sigma=\uparrow,\downarrow} \frac{(\pi k_B)^2}{12} D^\sigma(\varepsilon_F,V) \cdot T^2, \qquad (3)$$

where $D^\sigma(\varepsilon_F,V)$ is calculated electronic density of states for each spin $\sigma$ at Fermi level $\varepsilon_F$. Magnetic part with the mean-field approximation can be subtract from the localized spins model as

$$F_{mag}(T,V) = -k_B T \cdot \ln(\mu[T,V]/\mu_B + 1), \qquad (4)$$

where $\mu$ is local magnetic moment. For our free energy calculations we used the averaged local magnetic moment at 0 K and at the constant volume $V_0$ $\mu[T,V] \approx \mu[0,V_0]$. So, the heat capacity at constant volume $c_v$ can be found by differentiating $F_{tot}(T,V)$ with respect to the temperature:

$$c_\upsilon(T) = -T \frac{\partial^2 F_{tot}}{\partial^2 T}. \qquad (5)$$

Finally, under the Debye approximation the Debye temperature can be estimated by fitting the theoretical heat capacity $c_v$ to the Debye model.

**Results**

The total phonon DOS as well as the projected DOS for Fe and N atoms obtained by the DFPT are shown in Figure 1. For the acoustic part of DOS we can see major weight of Fe vibrations due to the large atomic masses difference between Fe and N. In addition, there is low Fe-N hybridization degree in the phonon DOS and almost no contribution of Fe to the optic part. By increasing external pressure the degree of hybridization between acoustic and optical modes decreas and lead to the phonon band gap opening. The shapes of the DOS curves for different pressures are rather similar. As it expected, the phonon frequencies increase with pressure. The effect induced due to the strength of covalent Fe-N bonds at high pressure.

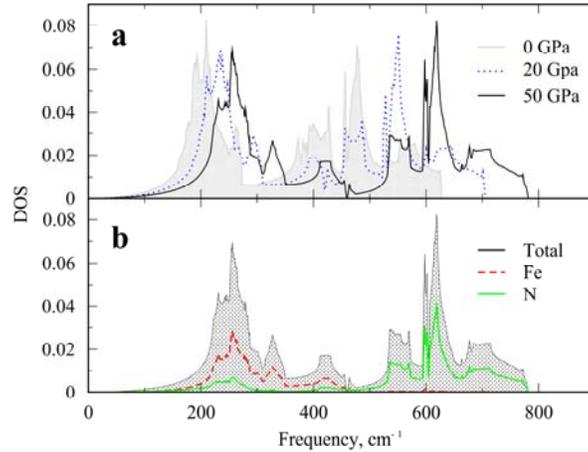

Figure 1. (a) Phonon density of states at 0 Gpa, 20 GPa and 50 GPa for ferromagnetic NiAs-FeN phase at zero temperature. (b) Total (per 4-atom unit cell) and partial phonon (per atom) density of states at 50 Gpa.

In Figure 2 we report curves of the Debye temperature and the heat capacity at constant volume versus temperature for the FM NiAs-FeN phase as obtained by QHA. The highest temperature is taken to be 500 K due to QHA limitation on accurate description of anharmonic effects at high temperatures. For the ground state at 0 K the value of Debye temperature is about 850 K and it decreases with temperature in low temperature region. The dependence of the Debye temperature on pressure is the same; pressure increasing tent to increase of the Debye temperature. Such behaviors bring us to the conclusion that suppression of the Debye temperature can be achieved by increasing of the atomic volume.

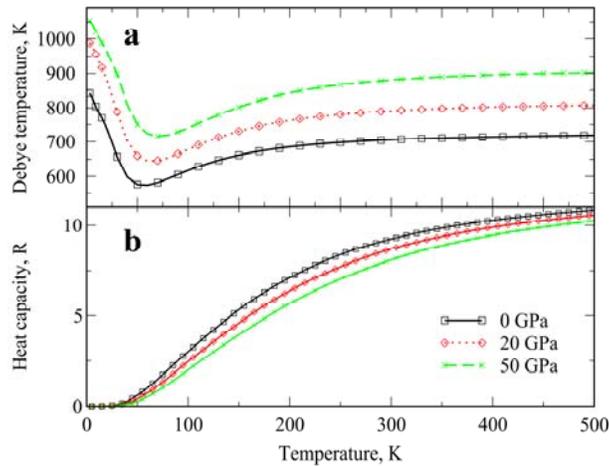

Figure 2. Temperature dependence of (a) the Debye temperature and (b) isochoric heat capacity per 4 atoms for different zero temperature pressure. Capacity provided in the units of gas constant R.

One $k_B$ will contributed to the isochoric heat capacity by fully exciting one vibrational branch. Comparing phonon DOS (Figure 1) and the isochoric heat capacity (Figure 2b), it can be seen that phonon modes with frequency ≤500 cm$^{-1}$ can fully activated at temperature < 250K. In this work, we will refer to isovolume heat capacity, since the isobaric heat capacity has similar trends as a function of temperature. Our results show that the heat capacity curves for different unit cell volumes are almost identical to each other. Just a minor slower increase of the heat capacity when increasing the temperature is found for higher pressure due to the phonon DOS expansion.

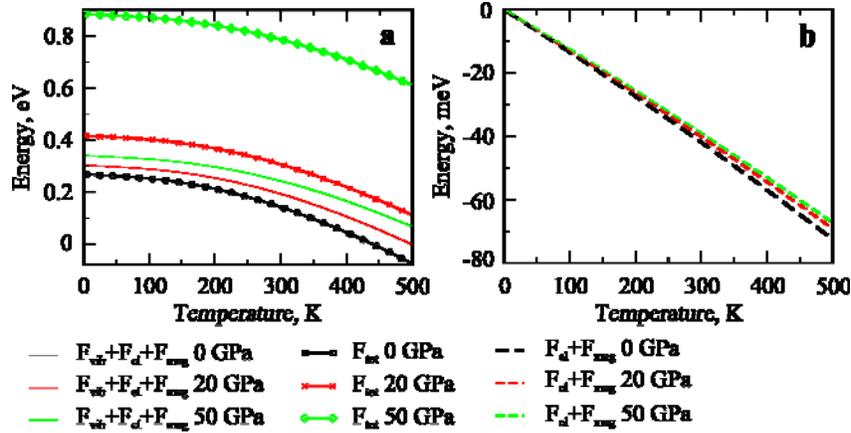

Figure 3. (a) Free energy and free energy without ground state contribution as a function of temperature for different fixed unit cell volumes of FM NiAs-FeN phase. (b) Electronic and magnetic contribution to free energy. Each unit cell volume maped to a specific hydrostatic pressure at zero temperature. Free energy at 0 GPa taken as a reference state, thus black curves at (a) are overlapped.

In Figure 3 we show the free energy vs temperature of the FM NiAs-FeN phase. The results shown in Figure 3b contain contribution due to the electronic thermal excitations and magnetic entropy in addition to the ground state energy and phonon part of the free energy. As one can see, the corrections from the electronic excitations and magnetic entropy are not large because of a low population near the Fermi level and relatively small local magnetic moment of FM NiAs-FeN phase. In the same time the shapes of the total free energy curves for every pressure are similar what provide the similitude of the heat capacity dependence.

**Summary**

The thermodynamic properties of FM NiAs-FeN phase is studied systematically by first-principles calculations based on DFPT scheme, including the vibrational, electronic and magnetic free energy contributions. The results of our DFT-GGA calculations indicate the pressure induced gap opening in the phonon DOS at 50 GPa. The ground state Debye temperature is found to be ≈ 850 K and ≈ 710 K for high temperature limit. Further studies of NiAs-FeN phase for extreme temperatures, which take anharmonic contributions for the thermodynamic properties and electronic correlations into account, will need to be undertaken. The calculations performed in this paper for high pressure FM NiAs-FeN phase provide important results for future theoretical and experimental investigations of the high concentration part of the Fe-N phase diagram and Earth's inner core structure.

**Acknowledgement**